# Modelling and simulation of complex systems: an approach based on multi-level agents

Alain-Jérôme Fougères[1]

[1]University of Franche-Comté / Computer Science Laboratory (LIFC)
Montbéliard, France

**Abstract**

A complex system is made up of many components with many interactions. So the design of systems such as simulation systems, cooperative systems or assistance systems includes a very accurate modelling of interactional and communicational levels. The agent-based approach provides an adapted abstraction level for this problem. After having studied the organizational context and communicative capacities of agent-based systems, to simulate the reorganization of a flexible manufacturing, to regulate an urban transport system, and to simulate an epidemic detection system, our thoughts on the interactional level were inspired by human-machine interface models, especially those in "cognitive engineering". To provide a general framework for agent-based complex systems modelling, we then proposed a scale of four behaviours that agents may adopt in their complex systems (reactive, routine, cognitive, and collective). To complete the description of multi-level agent models, which is the focus of this paper, we illustrate our modelling and discuss our ongoing work on each level.

***Keywords:*** *Agent-Based System, Agent Modelling, Agent Behaviour, Complex System, Multi-Level Agent*

## 1. Introduction

The main objective of our research over the last decade has been agent-based simulation and complex system (CS) modelling. After studying the agent-based systems (ABS) organizational context, to simulate the reorganization of flexible manufacturing and regulate an urban transport system, we focused on modelling agents with strong communication skills, which may be used as building elements for the design of assistance systems to CS users. Our thoughts on the interactional level were inspired by models from human-machine interfaces field, especially those of the "cognitive engineering" approach [6]. Then we suggested a first agent model and defined the communication model of these agents [10,11,29]. Next, we realized that the agent architecture could vary (multi-level architecture), so as to support the more or less cognitive tasks that agents perform. We then defined the different granularities of these agents by placing them on a scale of behaviours inspired by Ramussen's three-level scale [31] to describe human operator behaviour.

CS are "made of many components with many interactions" [34]. The CS design (cooperative systems, assistance systems, etc.) then includes a very specific modelling of interactional and communicational levels.

Moreover, according to Morin [22], CS designer "must have a method that allows to design the multiple points of view, and to move from one point of view to another."

In this sense, Wooldridge [37] and Jennings [19] have argued that agents are a new paradigm for CS engineering; they suggest a satisfactory response to three common techniques for reducing the software complexity: decomposition, abstraction and organization. Jennings [19] raises two hypotheses of adequacy and formation: (1) agent approach can significantly improve our ability to model, design and build complex and distributed software systems, (2) in addition to being able to design and build CS, agent approach is destined to become a major paradigm of software engineering.

From the IAD field [36], ABS, and before that the actor model [17], all withhold the basic principle of the knowledge and information distribution necessary to solve a problem on a set of interacting agents, capable of pursuing and achieving a common goal. An agent is an active, interactive and proactive entity of a system. It is often differentiated according to its cognitive (social metaphor) or reactive (biological metaphor) nature; we frequently define the granularity of agents according to their degree of knowledge and functional complexity.

A software agent, according to the Newell and Simon model [24], is an autonomous information processing system that means it is composed of reception and transmission devices, a processor and a memory (knowledge base). An agent-based system is a society of autonomous agents working together to reach a common goal from interaction, communication or transaction. For a first understanding of the agent paradigm, we can consider that an agent is a computer system located in an environment, in which it can act, possibly in interaction with others agents, and this in complete autonomy [19]. Autonomy is for us the main characteristic of an agent, relating to the object paradigm. It is realized by: (1) an independent computer process, (2) an individual memory (knowledge/data), and (3), an ability to interact (perception/reception, communication/action).

The paper is organized as follows. In the second section, the adequacy of the agent paradigm to model and simulate complex systems, and then a model of multi-level agents is suggested. In the third section, an illustration and a discussion of each of the four agent behaviour levels was drawn. Finally, in the last section, the conclusions of this research are presented.

## 2. Agents for Complex Systems Simulation

### 2.1. Agent Paradigm

There are many definitions of the agent paradigm [19, 37], supported by typology proposals, but new types of agents continue to emerge [35]. It is therefore difficult to establish a consensus. However, through these definitions we observe that three functions characterize agent activity: to perceive, decide, and act. An agent has its own knowledge. It acts in autonomy to reason and decides according to its objectives, its interactions with other agents in the system, and its environment perception. By extension, considering cognitive agents, experts of this domain generally agree on the following characteristics: intentionality, rationality, commitment, adaptability, and "intelligence". ABS are systems that allow distributing agents, communicating, autonomous, reactive, skilful, and finalized entities. They form intelligent solver networks, weakly bound, working together to solve problems beyond their individual capabilities and knowledge [18].

Among the suggested agent architectures with a cognitive orientation, the *BDI* model (Belief, Desire, and Intention) is best known [32]. It is built around three concepts inspired by human behaviour models: (1) beliefs, based on agent knowledge, (2) desires, corresponding to the knowledge that agent would express, and (3) intentions, or actions, that agents decide to do. Thus, a software agent, which has a desire, can transform it into intention when it knows it can achieve this; it is then left to act!

### 2.2. Agent Organisation and Interaction

Work in sociology or science organizations have always interested the ABS community, as a source of modelling [38]; the organizational configuration type scale by Mintzberg [21], from hierarchical structure (centralization) to the professional bureaucracy (decentralization), via the "Adhocracy" (mutual adjustment between groups), is one example.

The inherent problems to the partial agent knowledge in the pursuit of local goals or interlaced agents require the development of advanced coordination mechanisms [9]. An organization must allow ABS to behave as a coherent whole, to solve a problem uniquely. It controls and coordinates the interactions between agents of the system, and structures their activities.

Communication and interaction are interrelated. Following the definitions proposed for man-machine dialogue [5], it is possible to distinguish these two concepts as follows: *interaction* is an exchange between agents and their environment, this exchange depends on the intrinsic properties of the environment in which agents are active; agent *perception* may be passive by receiving messages/signals, or active, when it is the result of voluntary action; *communication* is an exchange between agents themselves, using a language. In addition, the concept of interaction is associated with the protocol and interaction pattern concepts, allowing us to specify how and with whom an agent can interact - engineering has been suggested [9]. Interaction specifications differentiate levels of interaction, such as micro or macro levels. As in the case of human communication, communicative interactions among agents can obviously be linked together and produce information exchanges or dialogues [2].

In most of ABS, the agent behaviour in interaction consists of three phases: (1) information reception from another agent, or the perception of a change in its environment, (2) interpretation of this event taking into account other agents, (3) sending a message or taking action in an environment modification. If interactions between agents are frequently communicative, they also involve action coordination, cooperation and negotiation [39].

### 2.3. Agent-Based Systems Design

The ABS design, often questioned from process or methodological perspectives [4], presumes that the designer proceeds with a local vision to respect the fact that each agent manages its own knowledge and actions (autonomy). The support languages of ABS design are numerous [3].

In [11] we proposed an ABS design method in four stages making extensive reference to *AUML* (Agent Unified Modelling Language) [1]: (1) making the use case diagrams (services provided by ABS); (2) for each case, sequence diagrams carry out the interactions (message exchanges and scheduling) between the agents involved in this reference case; (3) from the sequence diagrams, which identified agents, objects and their interactions, making the class diagram: objects are associated with classes, exchanged messages (service requests between objects) are translated into operations on classes, parameters associated with operations are translated into class attributes - it may be possible to complete this diagram by a collaboration diagram; (4) from the class diagram, defining the behaviour of each agent (class agent) with a state diagram or an activity diagram. The description of roles played by various cooperative agents is insufficient since the notion of role, is non-existent from *UML*, and isn't the subject of a few short extensions in *AUML* [1]. It focuses on collaboration diagrams and sequence diagrams.

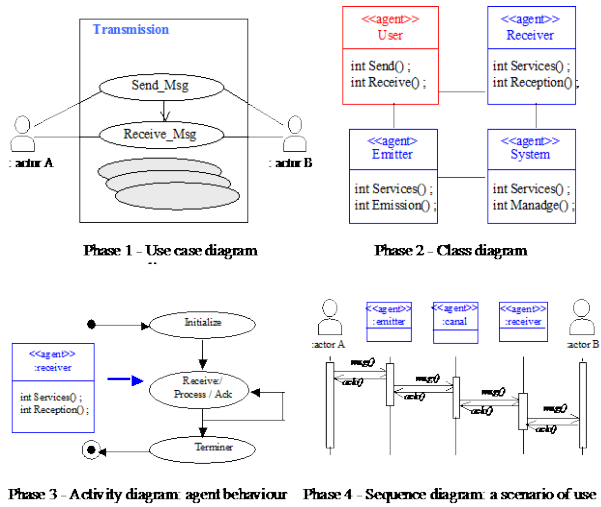

Fig. 1. *Agent design methodology, illustrated by a specification of "Message Transmission"*

### 2.4. Proposition of Agent-Based Systems Modelling

After studying the ABS organizational context, to simulate the reorganization of a flexible manufacturing system and the regulation of an urban transport system, we focused on the modelling of agents with strong communication skills, which may be used as building the foundation for the design of complex systems such as cooperative systems and user assistance systems. Our thoughts about this interactional level were fed by models from the man-machine interface field, especially those of the "cognitive engineering" approach [6]. We then proposed a variable granularity agent model based on Rasmussen's three level scale [10, 11, 28].

*2.4.1 Notations*

Following notations and definitions are used in this modelling section:

$A = \{\alpha_i\}$ is the agents finite set, $i \in I_A$, $I_A = \{1,2,...,q_A\}$;

$I = \{\iota_i\}$ is the interactions finite set defined for all agents, $i \in I_I$, $I_I = \{1,2,...,q_I\}$;

$P = \{\rho_i\}$ is the roles to be played by all agents finite set, $i \in I_P$, $I_P = \{1,2,...,q_P\}$;

$O = \{o_i\}$ is the agent organizations into communities finite set, $i \in I_O$, $I_O = \{1,2,...,q_O\}$;

$\Sigma = \{\sigma_i\}$ is the sates finite set of agent-based system, $i \in I_\Sigma$, $I_\Sigma = \{1,2,...,q_\Sigma\}$;

$\Sigma_{\alpha_i} \subseteq \Sigma$ is the states finite set of agent $\alpha_i$;

$\Sigma_{M_{\alpha_i}} \subseteq \Sigma$ is the states finite set of agent-based system that agent $\alpha_i$ knows;

$\Pi = \{\pi_i\}$ is the observations finite set, $i \in I_\Pi$, $I_\Pi = \{1,2,...,q_\Pi\}$;

$\Pi_{\alpha_i} \subseteq \Pi$ is the observations finite set that agent $\alpha_i$ can do;

$\Delta = \{\delta_i\}$ is the decision rules finite set, $i \in I_\Delta$, $I_\Delta = \{1,2,...,q_\Delta\}$;

$\Delta_{\alpha_i} \subseteq \Delta$ is the decision rules finite set that agent $\alpha_i$ can trigger;

$\Gamma = \{\gamma_i\}$ is the actions/reactions finite set of all agents, $i \in I_\Gamma$, $I_\Gamma = \{1,2,...,q_\Gamma\}$;

$\Gamma_{\alpha_i} \subseteq \Gamma$ is the actions finite set that agent $\alpha_i$ can process;

$\Gamma_{\Lambda_{\alpha_i}} \subseteq \Gamma$ is the specific communication acts finite set that agent $\alpha_i$ can process;

$\Omega = \{\omega_i\}$ is the interpretations finite set of all agents, $i \in I_\Omega$, $I_\Omega = \{1,2,...,q_\Omega\}$;

$\Omega_{\alpha_i} \subseteq \Omega$ is the finite set of interpretations of observations made by agent $\alpha_i$;

$K = \{\kappa_i\}$ is the knowledge finite set of, $i \in I_K$, $I_K = \{1,2,...,q_K\}$;

$K_{\alpha_i} \subseteq K$ is the knowledge finite set of agent $\alpha_i$, with $K_{\alpha_i} = P_{\alpha_i} \cup \Sigma_{\alpha_i} \cup \Sigma_{M_{\alpha_i}}$;

$E = \{\varepsilon_i\}$ is the events finite set, $i \in I_E$, $I_E = \{1,2,...,q_E\}$;

$E_{\alpha_i} \subseteq E$ is the events finite set that agent $\alpha_i$ can observe;

$X = \{\chi_i\}$ is the conditions finite set, $i \in I_X$, $I_X = \{1,2,...,q_X\}$;

$X_{\alpha_i} \in X$ is the conditions finite set associated to the internal states of agent $\alpha_i$;

$N = \{\nu_i\}$ is the configuration networks finite set of, $i \in I_N$, $I_N = \{1,2,...,q_N\}$;

$\Upsilon = \{\upsilon_i\}$ is the connexions between agents finite set, $i \in I_\Upsilon$, $I_\Upsilon = \{1,2,...,q_\Upsilon\}$;

$\Lambda = \{\lambda_i\}$ is the speech acts finite set, $i \in I_\Lambda$, $I_\Lambda = \{1,2,...,q_\Lambda\}$;

$H = \{\eta_i\}$ is the messages finite set, $i \in I_H$, $I_H = \{1,2,...q_H\}$;

$T = \{\tau_i\}$ is the type of messages finite set, $i \in I_T$, $I_T = \{1,2,...,q_T\}$;

$\Phi_{\Pi(\alpha_i)} : \Sigma \times \Sigma_{M_{\alpha_i}} \rightarrow \Pi_{\alpha_i}$ is the function of observations of agent $\alpha_i$;

$\Phi_{\Omega(\alpha_i)} : \Sigma_{M_{\alpha_i}} \times K_{\alpha_i} \to \Omega_{\alpha_i}$  is the function of interpretations of agent $\alpha_i$;

$\Phi_{\Delta(\alpha_i)} : \Pi_{\alpha_i} \times \Sigma_{\alpha_i} \to \Delta_{\alpha_i}$  is the function of decisions of agent $\alpha_i$;

$\Phi_{\Gamma(\alpha_i)} : \Delta_{\alpha_i} \times \Sigma \to \Gamma_{\alpha_i}$  is the function of actions of agent $\alpha_i$.

*2.4.2 Modelling*

The formal approach we follow to model and design CS is to define the modular architecture of agents, to define their model of interaction, communication and knowledge and to respect a rigorous methodology for acquiring expertise. Then, an ABS *M* is described by a 4-tuple (1):

$M = <A, I, P, O>$ (1)

where *A* is a set of agents, *I* is the set of interactions defined for agents of *A*, *P* is the set of roles to be played by agents of *A*, and *O* is the set of organizations of agents into communities.

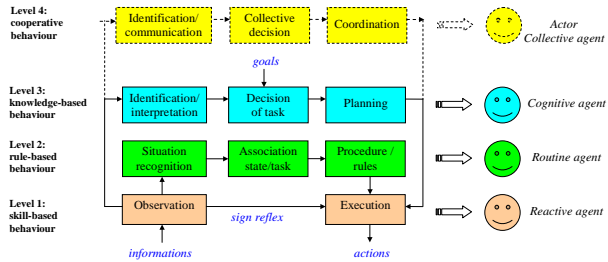

Fig. 2. *Variable agent behaviour, based on Rasmussen's model.*

Many agent structures known as "cognitive" are inspired by the cycle <*perceive, decide, act*> [23]. However, our agent model [10, 11] is rather inspired by Rasmussen's three-level operator [31]: 1) reflex-based behaviour, 2) rule-based behaviour, and 3) knowledge-based behaviour with interpretation, decision and plan (Figure 2). We interpreted this model as a model of process for agents. The latter are both cognitive and reactive. Moreover, they have behaviours adapted to the tasks they perform. We added one level at this scale to include behaviour based on cooperation. We call *actor* (or collective agent) a system of cooperative agents in which the behaviour is defined by collective decision tasks and collective coordination tasks [13].

Agents developed in our various projects, whose behaviour is illustrated in the following, can perform reflex actions (automatic), routine actions, and actions in new situations (creative or cooperative situations). These models are based on the sequential character of the human cognitive system (serialization in the symbolic level), not excluding a certain degree of parallelism in the processing of sensorimotor signals [34]. Thus, an agent $\alpha_i \in A$ can evolve on one of the first three levels of our scale (Figures 2 and 3); it is described by one of the following tuples (2, 3, 4):

$\alpha_i = <\Phi_{\Pi(\alpha_i)}, \Phi_{\Gamma(\alpha_i)}>$ (2)

$\alpha_i = <\Phi_{\Pi(\alpha_i)}, \Phi_{\Delta(\alpha_i)}, \Phi_{\Gamma(\alpha_i)}, K_{\alpha_i}>$ (3)

$\alpha_i = <\Phi_{\Pi(\alpha_i)}, \Phi_{\Omega(\alpha_i)}, \Phi_{\Delta(\alpha_i)}, \Phi_{\Gamma(\alpha_i)}, K_{\alpha_i}>$ (4)

where $\Phi_{\Pi(\alpha_i)}$ is the function of observations of agent $\alpha_i$; $\Phi_{\Omega(\alpha_i)}$ is the function of interpretations of agent $\alpha_i$; $\Phi_{\Delta(\alpha_i)}$ is the function of decisions of agent $\alpha_i$; $\Phi_{\Gamma(\alpha_i)}$ is the function of actions of agent $\alpha_i$; $K_{\alpha_i}$ is the finite set of knowledge of agent $\alpha_i$ - the knowledge contained in its memory, among which are the decision rules, the values of the domain, and the acquaintances and/or networks of affinities between agents, along with dynamic knowledge (observed events, internal states, etc.). The resource management associated with these various functions is provided by the set $M_G$ of managers: $M_G = \{M_H, M_\Gamma, M_K\}$, where $M_H$ is the messages manager, $M_\Gamma$ is the actions manager and $M_K$ is the knowledge-base manager (Figure 3).

The decision rules $\Delta_{\alpha_i}$ of the agent $\alpha_i$, gathered in its knowledge base, are described by a 3-tuple (5):

$\Delta_{\alpha_i} = <E_{\alpha_i}, X_{\alpha_i}, \Gamma_{\alpha_i}>$ (5)

where $E_{\alpha_i}$ is the set of events that agent $\alpha_i$ can observe, $X_{\alpha_i}$ is the set of conditions associated to the internal states of agent $\alpha_i$, and $\Gamma_{\alpha_i}$ is the set of actions that agent $\alpha_i$ can perform. For instance, let us consider the decision rule $\delta_1$ with: (1) $\varepsilon_1 := <inform, \alpha_{f_i}, \alpha_{r_k}, t = 2, V>$ ; (2) $\chi_1 := <V = sup(0.4)>$ ; (3) $\gamma_1 := <diffuse, \alpha_{f_i}, F, t=2, V>$.

This rule means that: (1) depending on following event $\varepsilon_1$: function agent $\alpha_{f_i}$ ($\alpha_{f_i} \in F, F \subseteq A$) receives a message of type *t* whose value is equal to 2 (corresponding to the transmission of a value) by which a requirement agent $\alpha_{r_k}$ ($\alpha_{r_k} \in R, R \subseteq A$) informs $\alpha_{f_i}$ of its value *V*; (2) under condition $\chi_1$ "*V* must be greater than the threshold value 0.4"; (3) action $\gamma_1$ will then be triggered: agent $\alpha_{f_i}$ will communicate this information to all function agents of the set *F*. Actions of agent $\alpha_i$ are controlled and memorized by a manager $M_{\Gamma_{\alpha_i}}$.

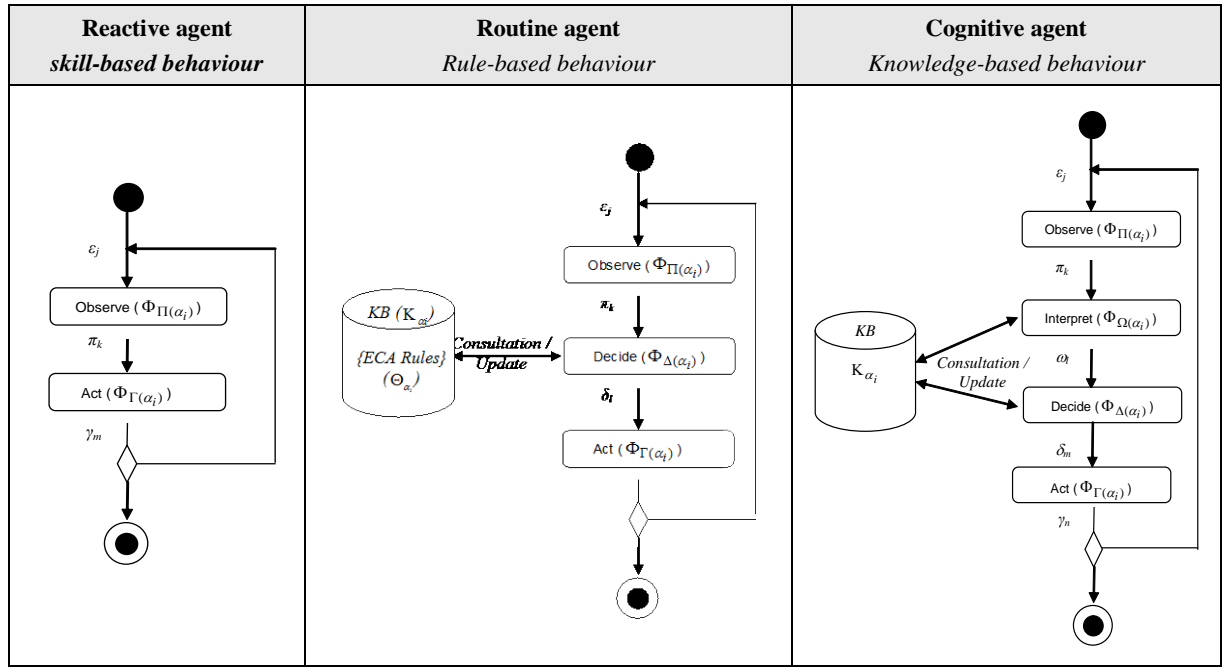

Fig. 3. *Activity diagrams for the behaviour of agents on the first three level of our scale*

**Agent communicational interactions.** Communication is the primary mechanism of interaction of an agent with its agent community. To communicate with other agents, an agent can exchange messages in syntax of an interaction language based on the concept of speech acts [33]. These information exchanges are controlled by a communication protocol in which a response is required for some speech acts (ask/[accept, refuse], inform/confirm, propose/[confirm, refuse], evaluate/[agree, disagree], etc.). We have compiled a lexicon of performative verbs we use in our applications ( $\Lambda$ = {*inform, diffuse, ask, reply, confirm...*}). The basic elements of this language (variables and primitives) are presented in Table 1. These speech acts are sufficient to enable agents to perceive the intention associated with the proposition content in a message. In the course of an interaction, an agent chooses its destination agent according to its intentions, the activity context and the state of its acquaintances.

A communication act $\lambda_{s,r}$ exchanged between two agents ( $\lambda_{s,r} \in \Gamma_{\Lambda_{\alpha_i}}$ ) is defined by a 5-tuple (6):

$$\lambda_{s,r} = <\lambda, \alpha_s, \alpha_r, \tau, \eta> \qquad (6)$$

where $\lambda \in \Lambda$ is a speech act denoted by a performative verb, $\alpha_s$ is the source agent of communication, $\alpha_r$ is the receiver agent, $\tau \in T$ is the type of message and $\eta \in H$ is the message itself, which can be an assertion, a question, a response, etc.

Table 1. Interaction language for cooperative agents

| Some elements of language | Significance |
|---|---|
| $\alpha, \varepsilon, \gamma, \eta, \beta, \tau$ | respectively are agent, event, action, message, speech act and type of message |
| inform($\alpha_s, \alpha_r, \tau, \eta$) | $\alpha_s$ sends to $\alpha_r$ the fuzzy message $\eta$ of type $\tau$ |
| diffuse($\alpha_s, [\alpha_l], \tau, \eta$) | $\alpha_s$ sends to the list $[\alpha_l]$ the fuzzy message $\eta$ of type $\tau$ |
| ask($\alpha_s, \alpha_r, \tau, \eta$) | $\alpha_s$ asks to $\alpha_r$ the fuzzy request $\eta$ of type $\tau$ |
| answer($\alpha_s, \alpha_r, \tau, \eta$) | $\alpha_s$ answers $\alpha_r$ the fuzzy message $\eta$ of type $\tau$ |
| confirm($\alpha_s, \alpha_r, \tau, \eta$) | $\alpha_s$ confirms to $\alpha_r$ that it is agree with fuzzy message $\eta$ of type $\tau$ |
| propose ($\alpha_s, \alpha_r, \tau, \eta$) | $\alpha_s$ proposes to $\alpha_r$ the proposition contained in the fuzzy message $\eta$ of type $\tau$ |
| against-propose ($\alpha_s, \alpha_r, \tau, \eta$) | $\alpha_s$ against-proposes to $\alpha_r$ the proposition contained in the fuzzy message $\eta$ of type $\tau$ |
| refuse($\alpha_s, \alpha_r, \tau, \eta$) | $\alpha_s$ refuses the proposition proposes by $\alpha_r$, contained in the fuzzy message $\eta$ of type $\tau$ |
| accept($\alpha_s, \alpha_r, \tau, \eta$) | $\alpha_s$ accepts the proposition proposes by $\alpha_r$, contained in the fuzzy message $\eta$ of type $\tau$ |
| order($\alpha_s, \alpha_r, \tau, \eta$) | $\alpha_s$ orders $\alpha_r$ to do the task contained in the fuzzy message $\eta$ of type $\tau$ |

**Coordination and organization of agents.** Interactions between agents of complex systems are not just communicational, they also involve cooperation and action coordination required to achieve the common goals of the agent system. Agent-oriented coordination models focus on the behaviour of agents in order to achieve a coordinated system. Initially organized in communities, agents are all involved in the same activity of production or problem solving. During activities, inter and intra-community networks of affinities between agents can emerge. Therefore, the coordination of agents and the self-organization of the communities are carried out by message exchange (mutual adjustment or emergence of networks of affinities, for examples).

**Agent knowledge.** An agent of a complex system has four kinds of knowledge: (1) domain knowledge, (2) functional skills to operate on states and values according to the processes defined in the complex system, (3) knowledge to control the agent's activity (decision rules), and (4) knowledge to interact with other agents (language and protocol of communication, acquaintances, networks of affinities, etc.). This knowledge is constantly evolving during tasks processed in the complex system, following the interactions between agents and the interventions of human users.

## 3. Illustration and Discussion

During the past decade, to validate the use of the proposed scale of agent behaviour (see Figure 2), we designed a set of agent-based systems and we also made numerous modelling. In the following we illustrate the agent modelling for each of our four-level scale with a representative application.

### 3.1. Level 1 (reactive agent): Agents to Design an Epidemic Simulation-Detection System

The main application for this level focused on the modelling and designs of simulation agents for an ABS called *SIMBADE* [10]. *SIMBADE* is an agent platform that aims to simulate and detect epidemics. This platform, designed in accordance with the French public health organization, allows simulating cases of disease (local or scattered) and regularly reporting diagnostic epidemic, from the messages exchanged by hierarchical agents of the detection system. *SIMBADE* combines complexity and clarity of presentation. Indeed, *SIMBAD* is composed of three subsystems (Figure 5.a): (1) an ABS for the simulation of epidemics, including agents from level 1 which interest us especially in this section; (2) an ABS for the detection of epidemics, including agents of Level 3; and (3) a system of decision support manager of medical knowledge to diagnose diseases and epidemics. Each agent of detection system has its own knowledge in relation to its role within the organization. Decision making is thus distributed. The simulation system is defined to control the two other systems and to understand what would be an autonomous system deployed on the public health system.

The ABS of epidemics simulation, related to artificial life systems, is composed of two types of agents at level 1: (1) <*contaminant*> agents, carriers of diseases known by the system, including reportable diseases, and (2) <*individual*> agents who consult their doctors if they are contaminated. An agent <*contaminant*> is introduced locally by an agent of simulation activating disease (influenza, meningitis, pertussis, listeriosis, etc.). This agent may, because of its proximity, infect healthy agent <*individual*>. It becomes bearer of the disease and may, in turn, move, contaminate other agents of its environment. Once infected, a <*Patient*> agent consults a <*Doctor*> agent from the doctors network (or medical community) (see scenario of Figure 5).

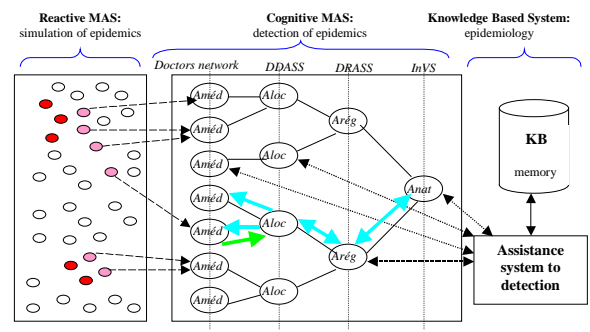

Fig. 4.  *The three-level architecture of SIMBADE*

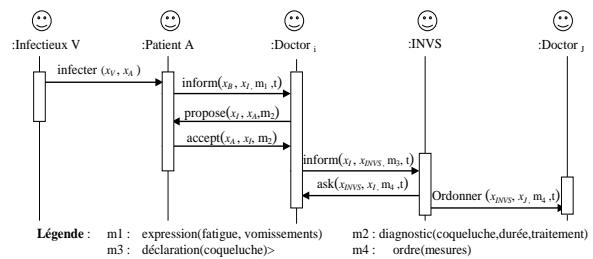

Fig. 5.  *Scenario describing the contamination of a <Patient> agent, leading to mandatory reporting.*

The agent behaviour, defined by the first level of our scale, is entirely satisfactory for this simulation activity. The reflex behaviour is inherently simple to model; we have not seen fit to reproduce it to validate the model at the first level.

### 3.2. Level 2 (routine agent): Agents to Design a Product Configuration System

The main application for this level focused on the modelling and designs of configuration agents for an ABS is called *APIC* [28]. *APIC* is an agent platform of collaborative design for an optimization of product integrated configuration. The configuration of product families is a collaborative and distributed process. Actors in the configuration access the *APIC* platform, using a specific μ-tool to their domain [20]. These μ-tools communicate with the platform through an interface agent. In *APIC*, any configuration item is an agent. Then, each specification (characteristic of product), function, solution (component of product) and constraint, is represented by an agent. Agents are organized in four communities that communicate heavily with each other and the interface agents (Figure 6): (1) community of <requirement> agents, (2) community of <function> agents, (3) community of <solution> agents, and (4) community of <constraint> agents. Each of these communities provides constraints on the process of configuration.

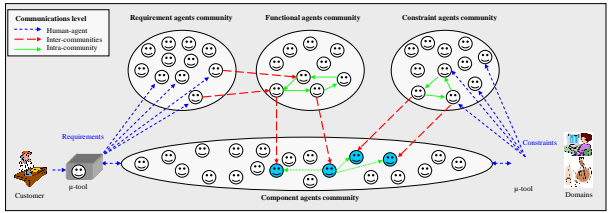

Fig. 6. Agent-based architecture of APIC platform: configuration/communication/cooperation levels

Note that APIC agents, defined by level 2 on our scale (using ECA rules to decide on their actions), were designed as fuzzy [16], in the sense that, not only the data they process are fuzzy (fuzzy knowledge), but their behaviours themselves are fuzzy. Indeed, the inter/intra-communities interactions are weighted by fuzzy values, this allows to weaken them, strengthen them, or even to inhibit them [29]. They operate in an environment defined by fuzzy variables and constraints of configuration.

Figure 7 shows a typical scenario of configuration highlighting the communicative context between level 2 agents of APIC. Consider the first exchanges. An agent <Ri> sends a message M1 of type T (fuzzy value that characterizes it in the Requirements network, for example) to an agent <Fj,>, then, according to the defined communication protocol, awaits an acknowledgment message. This acknowledgment message will allow agent <Ri> to know that agent <Fj> has received the message M1 and was able to handle it. For its part, agent <Fj> consults its ECA rules. One of them includes: *Event* [receiving a message of type T], *Conditions* (value > threshold, for example), and *Action* [diffusion to entire functions community]. Then, agent <Fj> broadcasts the received message, and when it received all acknowledgments of his own message, it can send the acknowledgment expected by agent <Ri>. And so on.

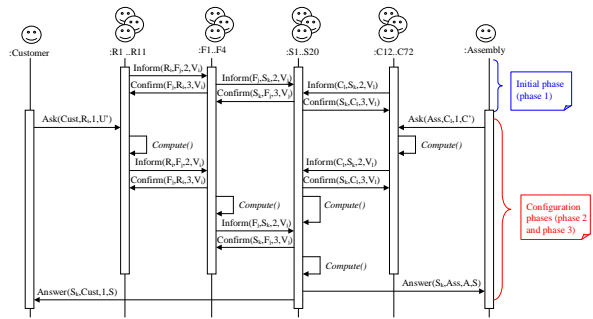

Fig. 7. Typical scenario for product configuration

The behaviour and knowledge (decision rules) of defined agents is entirely satisfactory for this configuration activity [29]. To further validate the model in this second level, we have recently reproduced for modelling simulation agents of a push service (information on mobility) for an on-demand transport system. Other cases of modelling in level 2 have been proposed [12, 13].

### 3.3. Level 3 (cognitive agent): Agents to Design an Assistance System for Student Projects Management

The main application for this level focused on the modelling and design of assistant agents for an ABS called *iPédagogique* [25]. *iPédagogique* is an agent-based platform that aims to track, manage, and evaluate students' projects. Management of students' project is a complex activity, cooperative and low-instrumented. After designing functionalities for the learning environment *iPédagogique*, it is natural that we were interested in the right level of assistance to provide for facilitating its use. Then, *iPédagogique* proposed an ideal testing environment for the design of an agent-based system in level 3 (high cognitive skills and specific knowledge to each agent). Modelling work has resulted in developing five assistant agents: (<aCourse>, <aPM>, <aUser>, <aForm>, <aTutorial>).

During the semester, the assistant of project management <aPM> simplifies communication, project management and monitoring: for the concerned teacher, assisting in the information broadcasting task, project registration, schedules, and group monitoring, etc.; for student projects groups, assisting in the project registration tasks, compliance schedules, sharing roles, delivery of documents, making appointments, phases constraint reminders, etc.; for tutor projects, assisting in the tasks of monitoring schedules for project groups, receiving documents, making appointments, etc.

The cooperative behaviour of agent <aPM> to a group of students comes in two modes: an intervention mode and an informational mode (a reminder during the connection of students, for example). As an illustration, Figure 8 shows the reaction of <aPM> following the delay of a project group during the analysis phase - this should normally result in the delivery of a requirements document to the tutor of the group. An interaction will be maintained with the student responsible for the phase in question. Then *<aPM>* observes proposals from students A and B and tutor, and decides to inform according to these defined *ECA* rules.

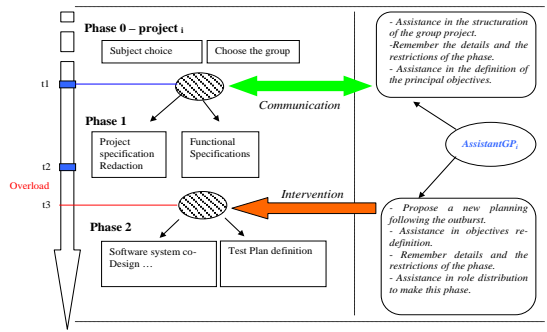

Fig. 8.    *<aPM> intervention following an overflow phase*

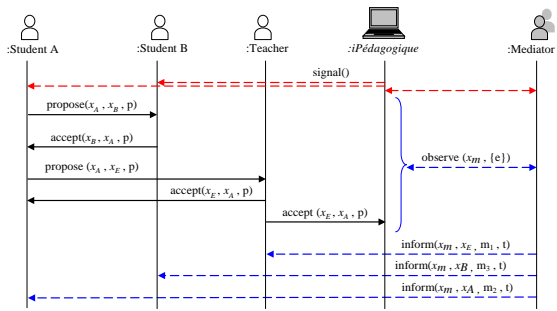

Fig. 9.    *Interactions between actors and Mediator during <aPM> intervention*

The behaviour of these assistant agents is entirely satisfactory for this activity of assistance to project management. To further validate the model in this third level, we have recently reproduced for modelling five design roles with cognitive agents (*<Prescriber>*, *<Legislator>*, *<Designer>*, *<Evaluator>*, and *<Observer>*), to realize a demonstrator with the aim of validating an advanced model of mechanical design [8]. Other cases of modelling have been proposed [10, 27].

### 3.3. Level 4 (collective agents): an Actor of Mediation for a Functional Analysis System

The main application for this level is focused on the modelling and designs of an agent-based system of cooperative mediation called *Mediator* [26, 27]. *Mediator* is an artificial actor integrated in collaborative systems. Its role is to assist users in their cooperative activities of design, project management, functional analysis, or software development. The *Mediator* is designed as a group of agents; each agent has specific skills of cooperation (communication, co-memorization, co-production, coordination, and control_process, that we called the *5-Co* [20]).

The *Mediator* is part of a group of human actors during a cooperative functional analysis activity, which means it is able to interact with them. For that it must perform cognitive tasks of observation, interpretation, decision and action.

The projection of the agents' skills in the *Mediator* activity pattern, allows to define that: the task "observe" will be performed by agents *<Observer>* specialized in the acquisition of cooperation information mediated by the cooperative system; the task "interpret" and "decide" will be performed by agents *<Knowledge>* and *<Control>*; monitoring of cooperative activity will be performed by the agent *<Monitoring>*; memorization by the agent *<Memorization>*, and the task "act" relating to the cooperation communication information (communication to act), will be performed by the agent *<Communication>*. Coordination is centralized at the agent *<Control>*. Figure 10 shows the agent architecture of *Mediator*. The cooperative system is artificially divided into two parts to facilitate understanding of the scheme. *<Observer>* agents are specialized by modes of cooperation, while *<Communication>* agent is unique (*ie*, performing tasks of similar nature).

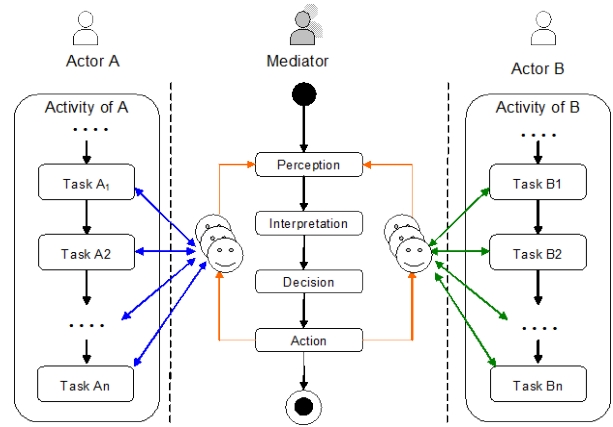

Fig. 10.   *Functional architecture of the Mediator*

To make explicit communication between *Mediator* and designers, we have made a strong case for cooperation: there can be interaction without a minimum level of cooperation, especially in the case of a mix of human and artificial actors. Communicative interactions are of two types: macro-interactions between the designers and the *Mediator*, and micro-interactions between the agents of *Mediator*. Figure 11 shows micro-interactions between the five agents comprising *Mediator*, resulting from the observation of a designer's proposal.

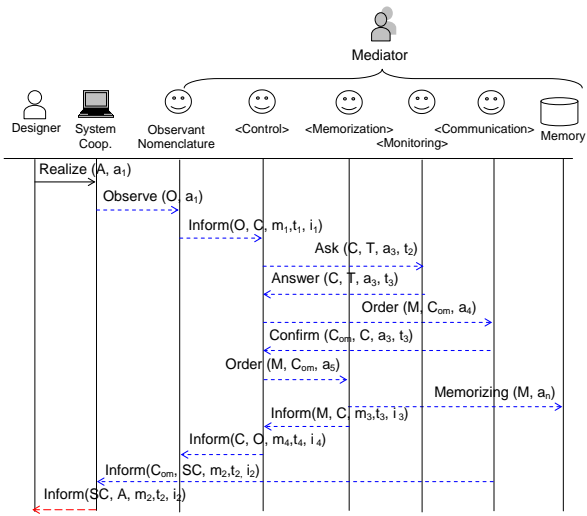

Fig. 11. *Interactions between agents of the Mediator, cooperative system and memory*

The cooperative behaviour of these agents, defined by the fourth level of our scale, is entirely satisfactory for this activity of cooperative mediation. To further validate the model in this fourth level, we have recently reproduced agents for modelling simulation for modelling design communities of *APIC*. These communities are seen as actors composed by the fuzzy cooperating agents previously described in the second level [28].

## 5. Conclusion and Perspectives

In this paper, we presented a generic framework for modelling agents defined for the modelling and design of complex systems (cooperative systems, assistance systems...). The correlated formal approach is to define a modular architecture for designing the various cognitive processes of agents, to respect a rigorous methodology to acquire expertise of each agent, to define their model of knowledge, and their interaction and communication patterns.

Agent activity in a complex system is very variable (from simple reflex reaction to complex cognitive decisions). So we have adapted the behaviour of agents modelled in our various applications. We then proposed a scale of behaviours which completes the scale proposed by Rasmussen. This scale consists of four levels we have presented in this article.

Recently, from the perspective of modelling communication and cooperation in CS, we have identified two levels of interaction situation: (1) micro-interaction situations, such as intra-*Mediator* cooperation (between agents of *Mediator*) or intra-community communication in *APIC* platform, for example; (2) macro-interaction situations, such as cooperation between the designers and the *Mediator*, or the communication between designers and requirement agents in *APIC* platform, for example. This extends our work on modelling of the interactions in complex systems with mixed prospects for communication between software agents and humans.

We are now working on a better understanding and implementation of level changes of agent behaviour during their activities [14]. The current extension of the *APIC* platform to other design tasks then offers an experimental context to test the evolution of our model.

## Références

**Alain-Jérôme Fougères** is Computer Engineer and PhD in Artificial Intelligence from University of Technology of Compiègne (UTC). He is currently member 1) of the Laboratory of Computer Science (LIFC) of the University of Franche-Comté, where he conducts research on mobility mediation for on demand transport systems; and 2) of the Laboratory of Mecatronics3M (M3M) at University of Technology of Belfort – Montbéliard (UTBM), where he conducts research on cooperation in design. His areas of interests and scientific contributions concern: the natural language processing, the knowledge representation, the agent-based system design (architecture, interactions, communication and co-operation problems). In recent years, his research has focused on the context of co-operative work (mediation of cooperation and context sharing).